\documentclass[aps,pof]{revtex4-1}
\pdfoutput=1

\usepackage{dcolumn}
\usepackage{bm}
\usepackage{amsmath}
\usepackage{amssymb}
\usepackage{mathtools}
\usepackage{graphicx}
\usepackage{natbib}
\usepackage{graphics,epsfig}
\usepackage{epstopdf}
\usepackage{float}
\usepackage{color}

\begin{document}

\preprint{AIP/123-QED}

\title{Standard logarithmic mean velocity distribution in a band-limited restricted nonlinear model of turbulent flow in a half-channel
}
\author{J.U. Bretheim}
 \email{jbretheim@jhu.edu}
 \affiliation{$^1$Department of Mechanical Engineering, Johns Hopkins University, 3400 North Charles Street, Baltimore,  MD 21218, USA} 
 \author{C. Meneveau}
 \email{meneveau@jhu.edu}
\affiliation{$^1$Department of Mechanical Engineering, Johns Hopkins University, 3400 North Charles Street, Baltimore, MD 21218, USA}
\author{D.F. Gayme}
 \email{gayme@jhu.edu} 
\affiliation{$^1$Department of Mechanical Engineering, Johns Hopkins University, 3400 North Charles Street, Baltimore, MD 21218, USA}

\date{\today}

\begin{abstract}
Numerical simulations of wall-turbulence using the restricted nonlinear (RNL) model generate realistic mean velocity profiles in plane Couette and channel flow at low Reynolds numbers. The results are less accurate at higher Re, and while a logarithmic region is observed, its von-K\'arm\'an constant is not consistent with the standard logarithmic law. In half-channel flow we show that limiting the streamwise-varying wavenumber support of RNL turbulence to one or few empirically determined modes improves its predictions considerably. In particular, the mean velocity profiles obtained with the band-limited RNL model follow standard logarithmic behavior for the higher Reynolds numbers in this study.
\end{abstract}

                             
\maketitle

One salient characteristic of wall-turbulence is the formation of well-defined, elongated structures aligned with the flow direction. These streamwise-coherent structures have inspired many numerical and experimental investigations, often with the goal of characterizing their role in the near-wall dynamics.\cite{Robinson91, Berkooz93, Henningson94, Hutchins07, Smits11} The structures have also motivated the development of streamwise-constant models of wall-bounded shear flows. One such model is the two-dimensional three-velocity-component (2D/3C) model, which has been studied in the context of both turbulent plane Couette flow\citep{Bobba04Thesis, Gayme10, Gayme11} and turbulent pipe flow. \citep{Bourg11} Under external stochastic forcing, the 2D/3C flow system develops elongated roll/streak structures consistent with wall-turbulence and its mean velocity profile transitions from a linear, laminar profile to the ``S-shaped'' profile characteristic of fully-developed turbulent plane Couette flow. \cite{Gayme10} While the 2D/3C model is nonlinear and captures the mechanism responsible for the blunting of the turbulent mean velocity profile, it does not include streamwise-varying velocity perturbations. It therefore requires continuous external excitation to generate a perturbation field and to maintain turbulence.\citep{Bobba04Thesis}

More recently, the restricted nonlinear (RNL) model has been proposed.  \cite{Thomas14,Constantinou14} The RNL model is similarly inspired by the prevalence of streamwise-coherent strucures in wall-turbulence.  However, it is more comprehensive than the 2D/3C model in that it describes the evolution of both a streamwise constant velocity field, which for the purpose of this paper will be referred to as the ``streamwise mean flow,'' as well as the evolution of a streamwise-varying perturbation field that interacts with the streamwise mean flow. The resulting coupling between the streamwise mean flow and the streamwise-varying perturbation field allows the RNL system to maintain turbulence through a self-sustaining cycle in which the streamwise mean flow is influenced by a perturbation field that is in turn regulated through interactions with the streamwise mean flow. \cite{Thomas14} The RNL model, which has been studied in the context of plane Couette flow \citep{Thomas14} and plane Poiseuille flow, \citep{Constantinou14} thus captures two mechanisms absent in the linearized NS system without the need for external stochastic forcing: the momentum transfer responsible for the turbulent mean velocity profile and the self-sustaining mechanism of turbulence.

In the RNL model, the  total velocity field, $\mathbf{u}_{\mathrm{T}}(x,y,z,t)$ (consisting of the respective streamwise, spanwise, and wall-normal components $(u_{\mathrm{T}},v_{\mathrm{T}},w_{\mathrm{T}})$ with $z$ as the wall-normal direction) is decomposed as 
$\mathbf{u}_{T} =  \mathbf{U}+\mathbf{u}$. Here $\mathbf{U}(y,z,t)=\langle \mathbf{u}_{\mathrm{T}}\rangle$ is the time-dependent streamwise-constant mean velocity, 
and $\mathbf{u}(x,y,z,t)$ is the streamwise-varying ``perturbation velocity." Here, angle brackets $\left\langle \ \right\rangle$ denote a streamwise-averaged quantity, averaged over the streamwise extent of the spatial domain
(i.e., the $k_x=0$ mode in a Fourier representation).  The RNL model dynamics can then be written as the following system of equations:
\begin{subequations}
\label{eq:whole}
\begin{equation}
\frac{\partial \mathbf{U}}{\partial t} + \mathbf{U} \cdot \nabla \mathbf{U} + \nabla \mathbf{P} / \rho - \nu \nabla^2 \mathbf{U} = - \left\langle \mathbf{u} \cdot \nabla \mathbf{u} \right\rangle + \partial_x p_{\infty} \ \bf{\hat{i}}, \quad \nabla \cdot \mathbf{U} = 0 \label{subeq:1}
\end{equation}
\begin{equation}
\frac{\partial \mathbf{u}}{\partial t} + \mathbf{u} \cdot \nabla \mathbf{U} + \mathbf{U} \cdot \nabla \mathbf{u} + \nabla \mathbf{p} / \rho - \nu \nabla^2 \mathbf{u} = 0, \quad \nabla \cdot \mathbf{u} = 0 \label{subeq:2}
\end{equation}
\end{subequations}
with density $\rho$, kinematic viscosity $\nu$, and constant pressure gradient forcing $\partial_x p_{\infty}$ in the streamwise direction. This system differs from the full incompressible NS equations (decomposed in this manner) only in the omission of the term $\left\langle \mathbf{u} \cdot \nabla \mathbf{u} \right\rangle - \mathbf{u} \cdot \nabla \mathbf{u}$ from the right-hand side of the evolution equation for the streamwise-varying perturbation field  \eqref{subeq:2}.

Prior RNL simulations have shown that most of the streamwise-varying modes decay in time, leaving only a modest number of streamwise-varying modes interacting with the $k_x=0$ mode (the precise number and nature of these modes depends on Reynolds number and channel size). \cite{Constantinou14, Thomas14} Here $k_x$ refers to the non-dimensional wavenumber $k_x = \delta k'_x = \delta 2\pi / L_x \times (n)$, where $n$ is a nonnegative integer. We explore the behavior of the RNL system in a half-channel configuration where we pre-select a set of streamwise-varying wavenumbers ($k_x\neq0$) and limit the RNL dynamics to this set of modes interacting with the streamwise mean flow (i.e., the flow associated with $k_x=0$). In this context, we investigate whether the RNL system can predict realistic logarithmic mean velocity distributions and characterize its behavior as we vary the set of $k_x\neq0$ modes supporting the ``band-limited'' RNL turbulence. 
 
We simulate the RNL system by restricting the dynamics in an existing direct numerical simulation (DNS) code. The code employs a pseudospectral discretization in the streamwise ($x$) and spanwise ($y$) directions along with a centered second-order finite-difference scheme in the wall-normal ($z$) direction. Time integration is achieved with a second-order Adams-Bashforth method. No-slip and stress-free boundary conditions are imposed at the bottom and top walls, respectively, with periodic conditions in the horizontal directions. The 3/2 rule is applied for dealiasing. All simulations are conducted in a half-channel box of size $[L_x, L_y, L_z]/\delta = [4\pi, 2\pi, 1]$, where $\delta$ is the half-channel's height. We use uniform mesh-spacing in all coordinate directions and have cross-plane resolution of $[\Delta y^+, \Delta z^+] \approx [7.0, 1.0]$ for all Reynolds number values considered herein. The half-channel box size and cross-plane resolution are thus comparable to those of the DNS calculated by Moser et al. \cite{Moser99} As usual, the superscript $+$ indicates scaling by the inner variables of friction velocity and viscous lengthscale ($u_{\tau} = \sqrt{\tau_{w}/\rho} = \sqrt{ -(\delta/\rho) \ \partial_x p_{\infty}}$ and $\delta_{\nu} = \nu / u_{\tau}$, respectively). The wall shear stress is $\tau_w$ and the friction Reynolds number is Re$_{\tau} = u_{\tau} \delta/\nu$.

As previously mentioned, the RNL system naturally supports fewer streamwise modes ($k_x$) than the NS system.\citep{Constantinou14,Thomas14} Constantinou et al. \citep{Constantinou14} reported that in a full-channel configuration at Re$_{\tau} = 950$ (with $L_x=\pi \delta$), the RNL system sustained the six lowest streamwise-varying wavenumbers ($k_x = 2,4,..,12$), in addition to the streamwise mean flow ($k_x = 0$). The energy of the seventh and higher wavenumbers ($k_x \geq 14$) decayed asymptotically to zero.  

\begin{figure}
\includegraphics[scale=.73]{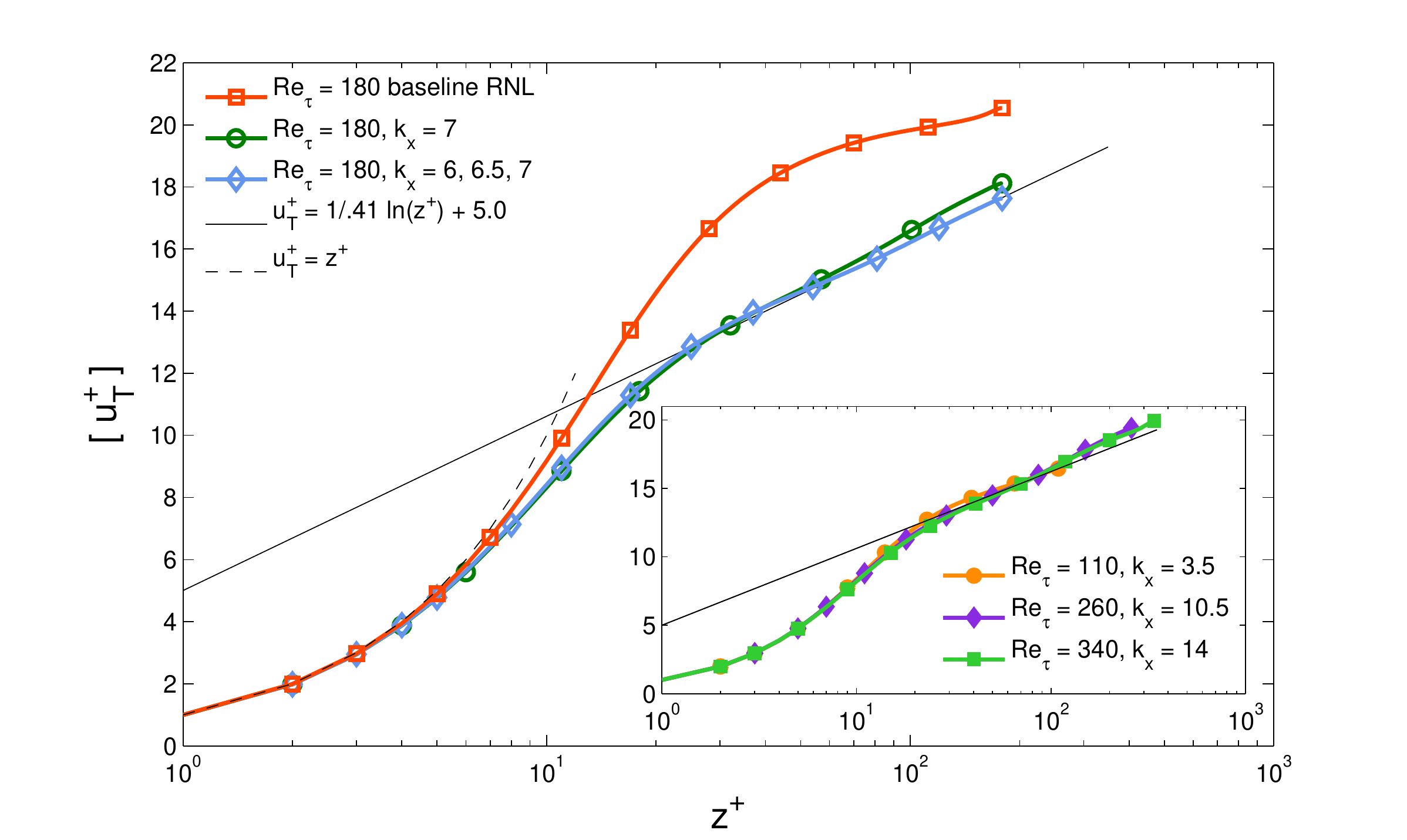}   
\caption{\label{fig:mvps} Mean streamwise velocity profiles for various simulated cases. The baseline RNL dynamics (red square markers in outer plot) is unconstrained, while all other cases have perturbation dynamics limited to the wavenumber(s) specified. Note: plot markers are sparse for data presentation purposes and do not indicate grid resolution.}
\end{figure}

Our simulations of  RNL turbulence  in the half-channel at  Re$_{\tau} = 180$ recover results similar to those of Constantinou et al.,\citep{Constantinou14} with the RNL turbulence naturally sustaining the five lowest streamwise-varying wavenumbers ($k_x = 0.5,1,1.5,2,2.5$) and the streamwise mean flow ($k_x=0$). The corresponding mean velocity distribution is shown in Figure \ref{fig:mvps} with open square markers. As is apparent, the baseline RNL system overpredicts the mean streamwise velocity for $z^+>10$ and produces an approximately logarithmic region above $z^+ \sim 30$ with slope and intercept that are, however, not consistent with the well-known values for $\kappa \sim 0.41$ and $B \sim 5$. This is consistent with the results of  Constantinou et al.,\citep{Constantinou14} who reported the existence of a logarithmic law in full-channel RNL turbulence with $\kappa = 0.71, B = 11.1$ at Re$_{\tau} = 350$ and $\kappa = 0.77, B = 14.0$ at Re$_{\tau} = 950$.

\begin{figure}
\includegraphics[scale=.45]{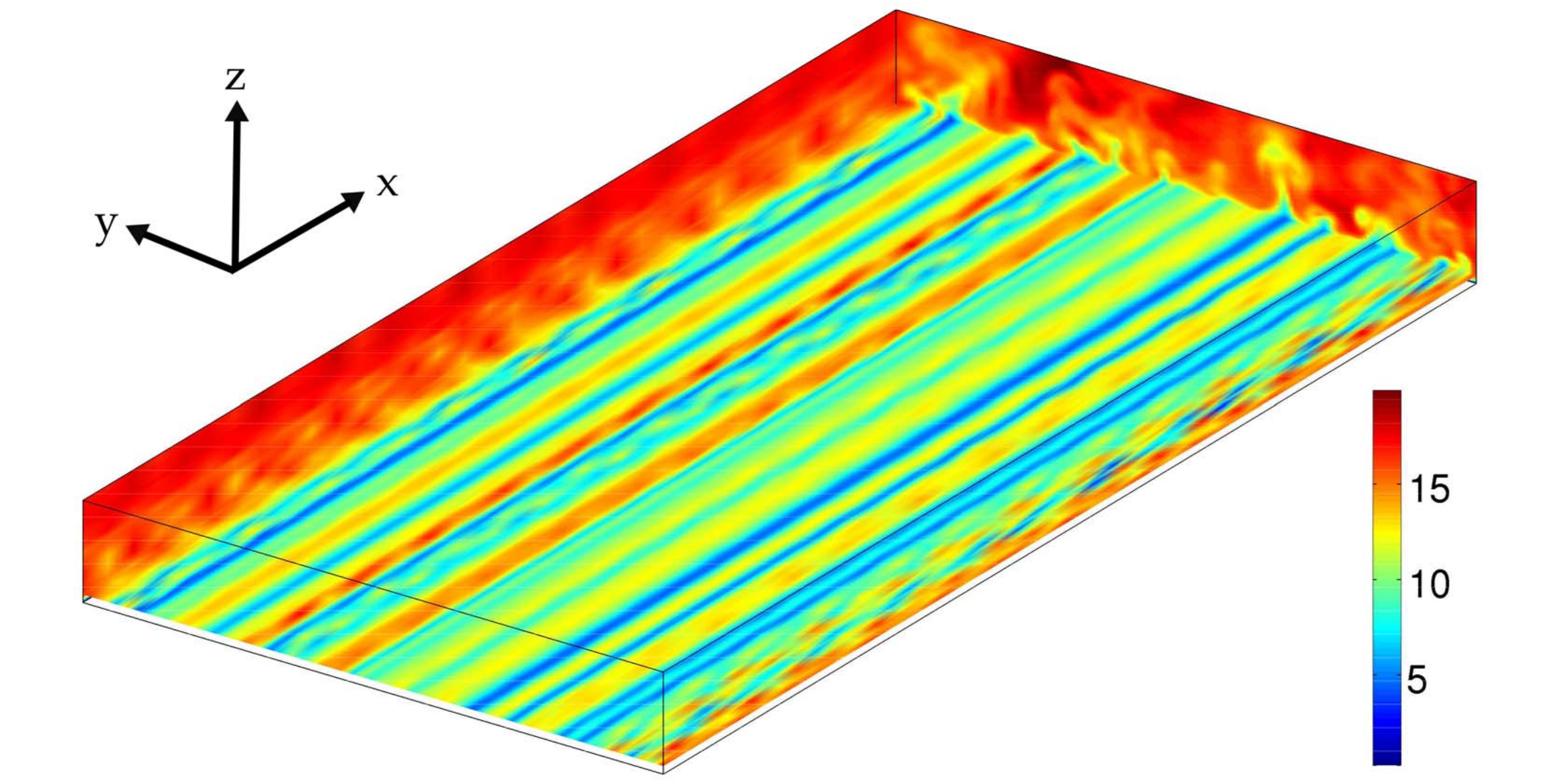}
\caption{\label{fig:box} Plane snapshots of streamwise velocity $u_{\mathrm{T}}$ in a RNL half-channel simulation at Re$_{\tau} = 180$. In this case, the streamwise dynamics is limited to the set of wavenumbers $k_x = [0, 6, 6.5, 7]$ in a box of size $[L_x, L_y, L_z]/\delta = [4\pi, 2\pi, 1]$. The horizontal plane is taken at height $z^{+} = 15$.}
\label{fig-snapshot}
\end{figure}

These results raise the question of whether keeping a different set of streamwise-varying wavenumber modes for the perturbation field could lead to more accurate interaction dynamics and therefore produce more realistic mean velocity distributions. Specifically, we consider an RNL system that is band-limited to a reduced set of streamwise-varying wavenumbers. To clearly distinguish the two types of RNL systems discussed herein, we will hereafter refer to the RNL system without any mode-limiting as the baseline RNL system. First, we experiment with various choices for supporting wavenumbers. For a set of three wavenumbers $k_x=6$, 6.5 and 7, the resulting mean velocity profile is shown in Fig. 
\ref{fig:mvps} with open diamond markers. This figure shows that the band-limited system generates profiles that closely match the standard logarithmic law.  In order to provide a qualitative view of the flow, a snapshot of the streamwise velocity is shown in Figure \ref{fig:box}.   The cross-plane structure displays realistic vortical structures while the band-limited nature of the streamwise-varying perturbations and the associated restriction to a particular set of streamwise wavelengths is clearly visible. These results demonstrate that the mean velocity profile obtained from simulations of the RNL system that are band-limited to only three streamwise-varying wavenumbers ($k_x = 6, 6.5, 7$) yields an accurate mean velocity profile. The profile exhibits a logarithmic region with standard values of $\kappa=0.41$ and $B=5.0$. 

The preceding results are achieved by constraining the dynamics of the perturbation field to a limited set of streamwise-varying modes, which essentially forces the flow to exist on some set of wavenumbers different from the set that naturally arises under the baseline RNL dynamics.  In previous studies the RNL system has also been shown to sustain turbulence even in the case of a single mode interacting with the streamwise mean flow. \cite{GaymeAPS} Figure \ref{fig:mvps} shows the mean turbulent velocity profile from such a case when the active streamwise-varying mode is  $k_x=7$ (open circle markers). These results show good agreement with the standard logarithmic law up to $z/\delta \sim 0.4$.  Thus the RNL system not only maintains turbulence when further limited to only one perturbation wavenumber but also yields an accurate mean velocity profile through interactions of $k_x = 0$ and $k_x=7$.  

Next, we examine a range of Reynolds numbers. In order to simplify the testing as much as possible, we consider a perturbation field with only a single streamwise-varying mode. In order to decide which one is retained in the model,  we simulate the RNL system at each Reynolds number for a range of single $k_x \neq 0$ wavenumbers. For each case, we quantify deviations between the model and the standard log law by comparing a measure of the integrated velocity profile with well-known empirical correlations for channel flow. A natural metric for this comparison is the skin-friction coefficient
\begin{equation}
c_f =  2 \tau_{w} / \left( \rho u_{0}^2 \right). \label{eq:c_f}
\end{equation}
In order to avoid relying on data at a single point (centerline velocity, $u_0$) we calculate the bulk average velocity, $\ u_{b} = \frac{1}{\delta} \int_{0}^{\delta} \left[ u \right] \mathrm{d}z$ from the simulations (here and in what follows, square brackets $\left[ \ \right]$ denote streamwise-, spanwise-, and time-averaged mean quantities). We relate  the bulk velocity to the centerline velocity using the empirical relation  $u_{0} \approx u_{\tau}/\kappa + u_{b}$. \cite{Pope} By calculating the skin-friction in this way, we obtain a single value that characterizes the entire mean velocity profile. We then compare the band-limited RNL system's $c_f$ to the following empirical correlation for $c_f$: \cite{Pope}   
\begin{equation}
\sqrt{\frac{c_f}{2}}=\frac{u_{0}}{u_\tau} = \frac{1}{\kappa}  \ln Re_\tau + B + B_1 \label{eq:u_0}
\end{equation}
with $\kappa = 0.41$, $B = 5.0$, and $B_1 = 0.2$.\cite{Pope} (We note that these values and the empirical correlation in Eq. \eqref{eq:u_0} are for a full and not a half-channel, but the differences are expected to be acceptably small for present purposes). We refer to the single streamwise-varying wavenumber which yields a $c_f$ value with the smallest absolute error relative to  Eq. \eqref{eq:u_0}  as the ``optimal" wavenumber.

\begin{figure}[htbp]
\begin{center}
\includegraphics[width=0.49\linewidth]{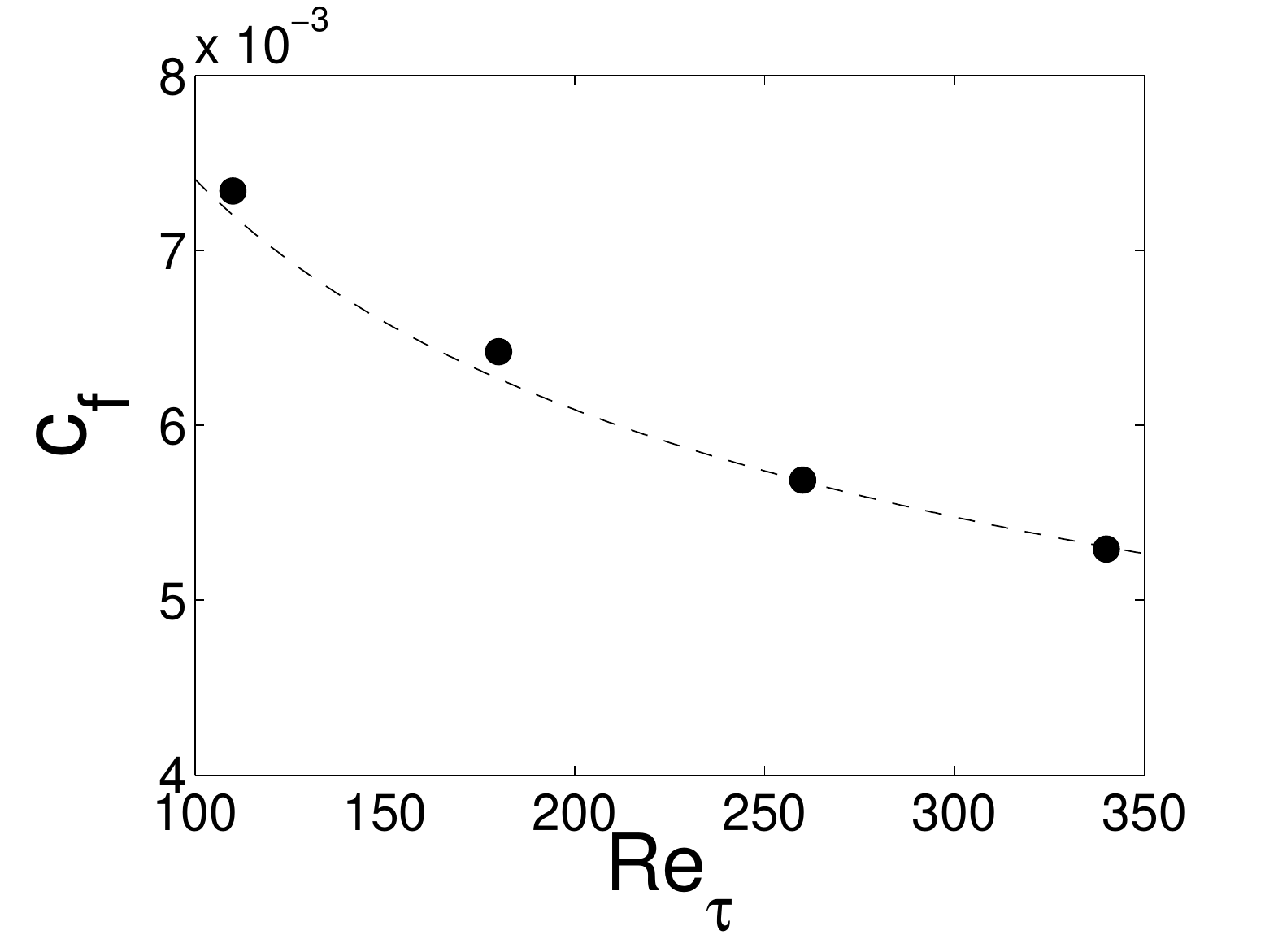}
\includegraphics[width=0.49\textwidth]{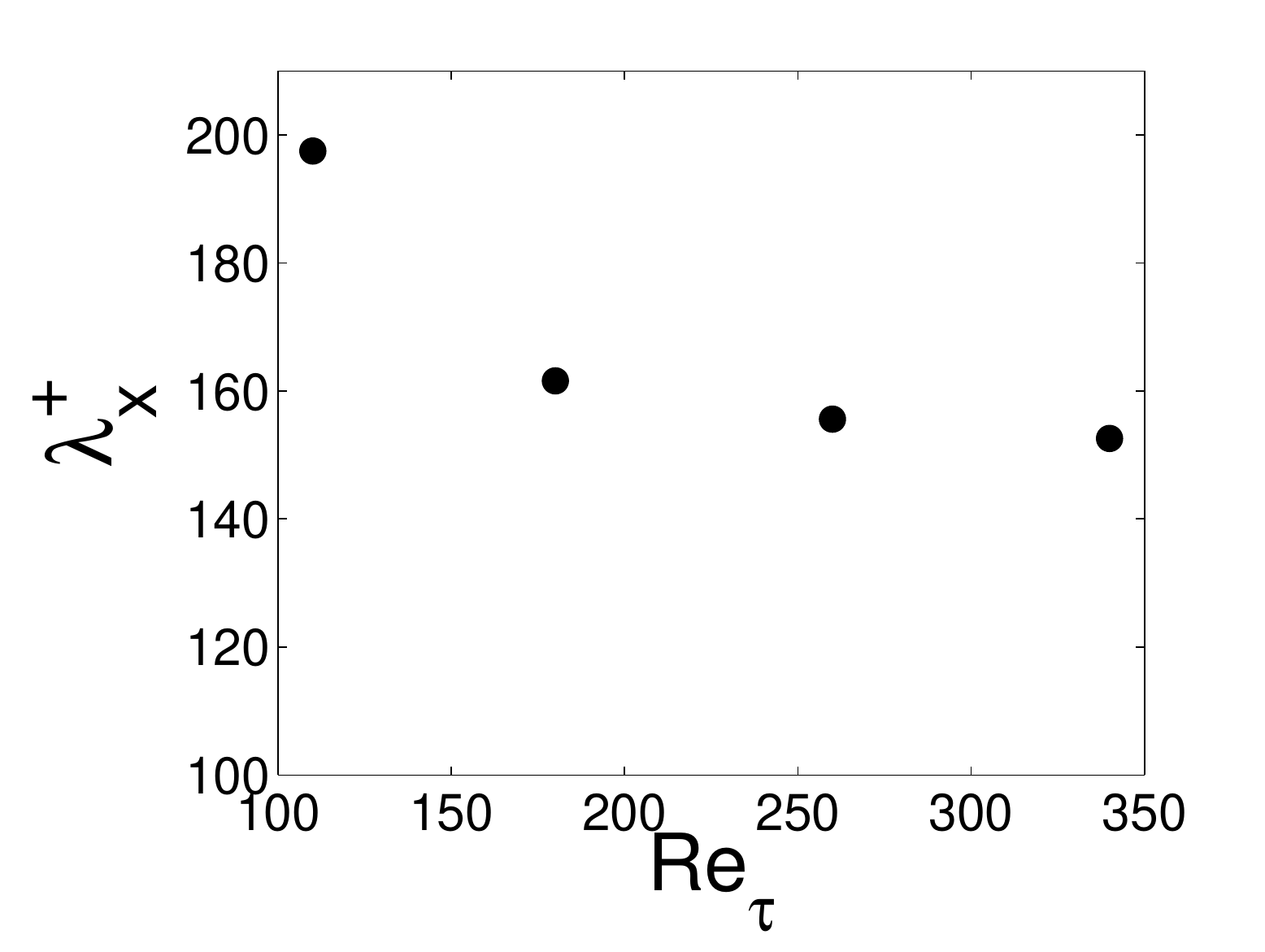}\\
(a) \hspace{0.45\linewidth} (b)
\end{center}
\caption{(a) Skin-friction coefficient $c_f$ as a function of friction Reynolds number. The dashed line is the standard empirical correlation for $c_f$ based on obtaining $u_0$ from equation \eqref{eq:u_0}. Circles are the band-limited RNL results run at the optimal $k_x$. (b) The optimal mode's corresponding streamwise wavelength, $\lambda_x = 2\pi/k_x$, as a function of friction Reynolds number (in inner units).}
\label{fig:c_f_lambda}
\end{figure}

Figure \ref{fig:c_f_lambda} shows the $c_f$ values obtained for the optimal streamwise wavenumber in panel (a) and the trend in the corresponding streamwise wavelength as a function of friction Reynolds number in panel (b). In initial tests, it was expected that perhaps $\lambda_x$ would tend to a certain fraction of $\delta$ (outer scaling), but the results appear to show inner scaling instead, tending to a wavelength of about 150 viscous units.  Further simulations at higher Reynolds numbers are required in order to establish whether this scaling can be maintained for arbitrarily high Reynolds numbers.  

\begin{figure}[htbp]
\begin{center}
\includegraphics[width=0.475\linewidth,clip]{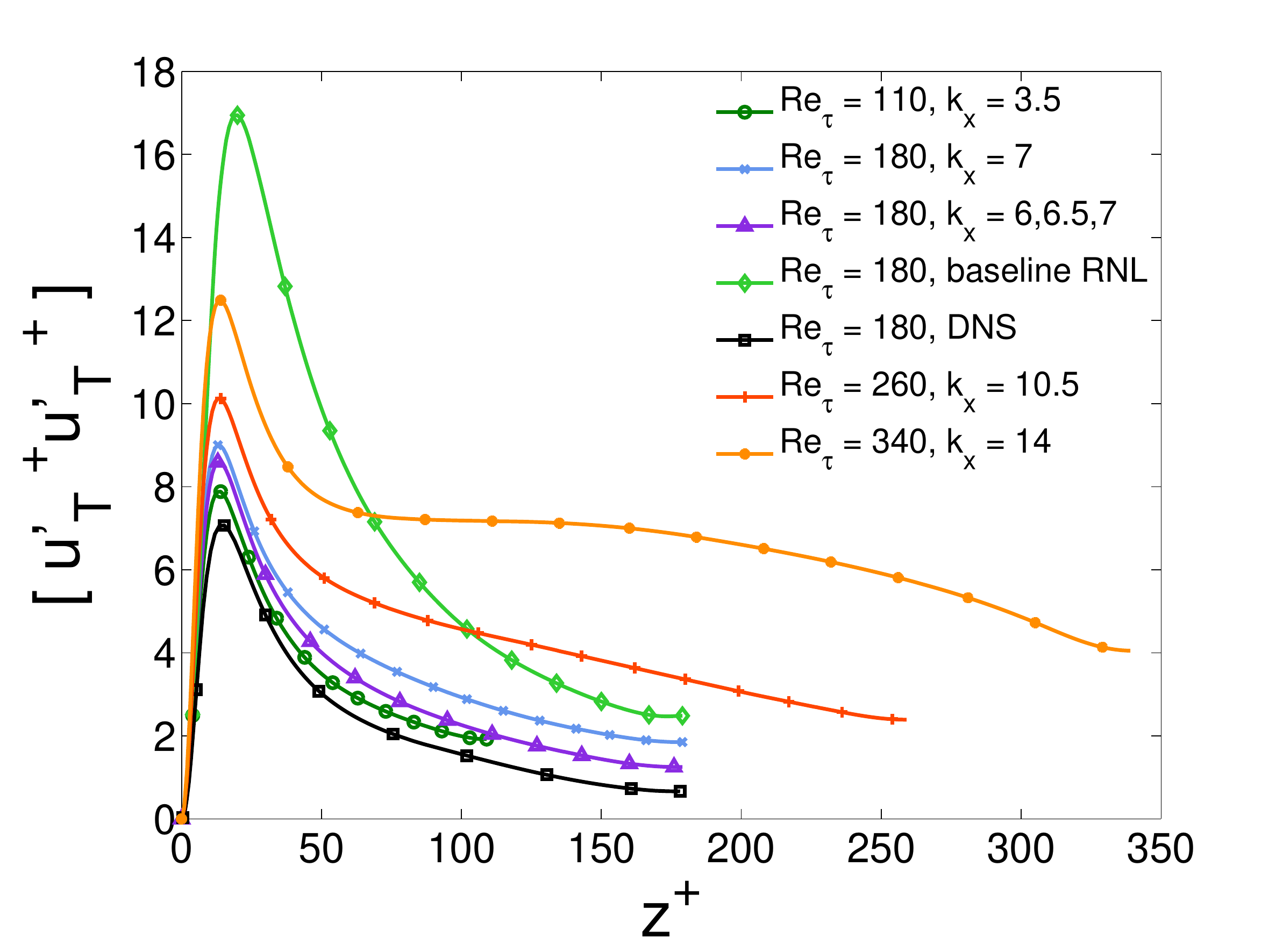} 
\includegraphics[width=0.475\linewidth,clip]{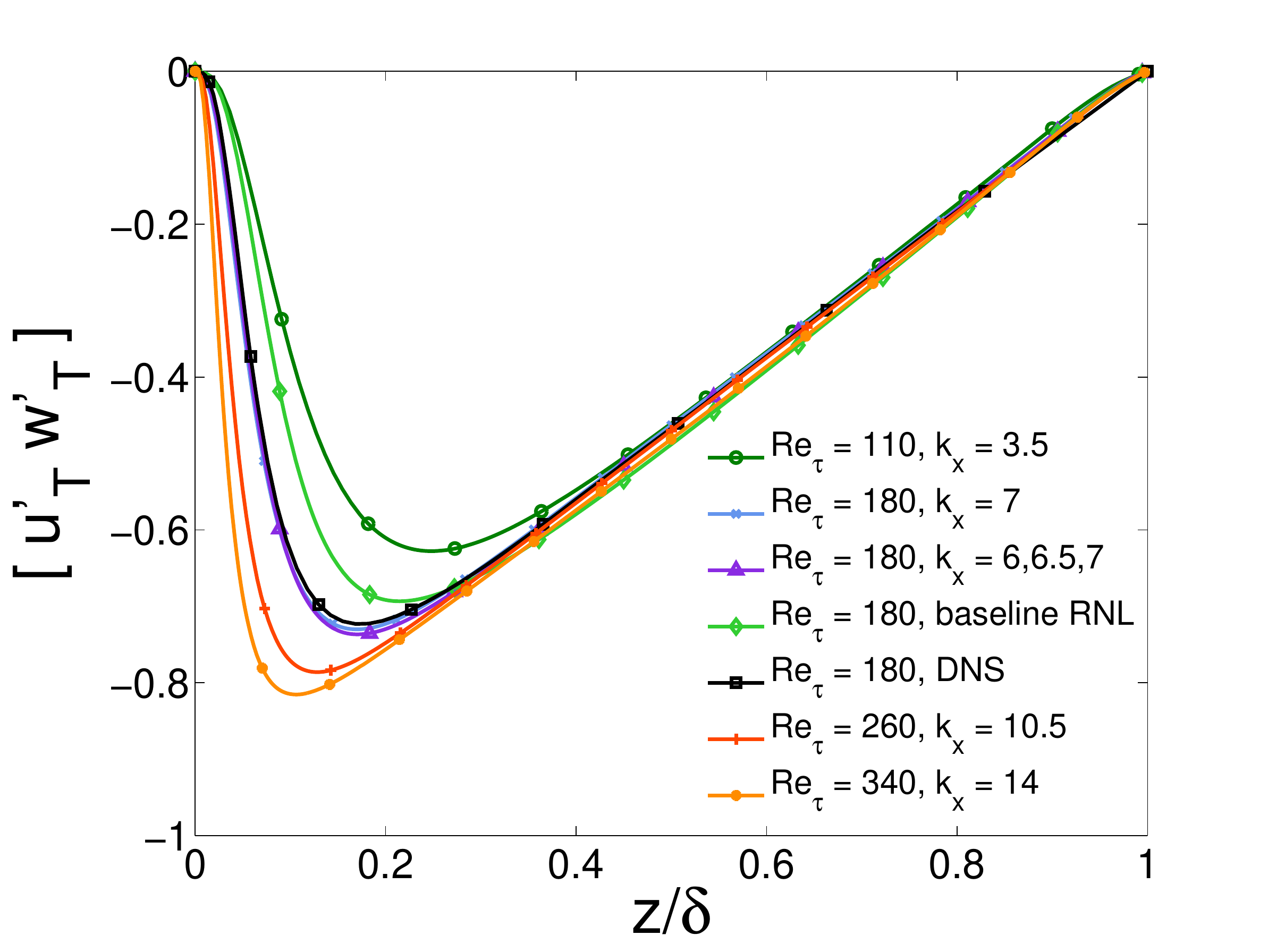}\\ 
(a) \hspace{0.45\linewidth} (b)
\end{center}
\caption{Profiles of (a) streamwise normal Reynolds stress (inner units) and (b) Reynolds shear stress (outer units). The superscript prime $^\prime$ indicates departure from the time-averaged value. The DNS values are from the simulation of Moser et al.\citep{Moser99}}
\label{fig:rs}
\end{figure}

Next, second-order statistics are considered. Figure \ref{fig:rs}(a) shows that the baseline RNL model overpredicts the normal Reynolds stress in the streamwise direction at Re$_{\tau} = 180$, with a peak value of $\approx$ 17 occurring at $z^+ \approx 20$. This differs from the Navier-Stokes DNS system's peak value of $\approx 7.06$ at $z^+ \approx 15.28$ (we compare with the simulation of Moser et al.\citep{Moser99}). By altering the streamwise wavenumbers supporting the RNL turbulence, however, we obtain improved predictions. When the RNL system's dynamics is constrained to interactions between the streamwise mean flow ($k_x=0$) and a single streamwise-varying wavenumber ($k_x=7$), the peak normal Reynolds stress in the streamwise direction is reduced to $\approx 9$ and occurs at $z^+ \approx 13$ for the Re$_{\tau}=180$ case. Permitting interactions with two additional wavenumbers, $k_x=6$ and $k_x=6.5$, yields a slight reduction to about $8.6$ for the peak value, occurring at the same wall-normal location. In general, the peak values of the streamwise components of the normal Reynolds stresses for the single streamwise-varying wavenumber cases increase with increasing Re$_{\tau}$. The wall-normal locations of these peak values change slightly to $z^+ \approx 14$ for both Re$_{\tau} =$ 260 and 340. As expected from overall momentum conservation, the Reynolds shear stress profiles, shown in figure \ref{fig:rs}(b), are quite realistic (since the mean velocity and hence the viscous stress distributions are realistic). Good agreement is obtained between the various RNL model cases at Re$_{\tau} = 180$ and the DNS of Moser et al.\cite{Moser99} at the same friction Reynolds number. 

\begin{figure}[htbp]
\begin{center}
\includegraphics[width=0.49\linewidth]{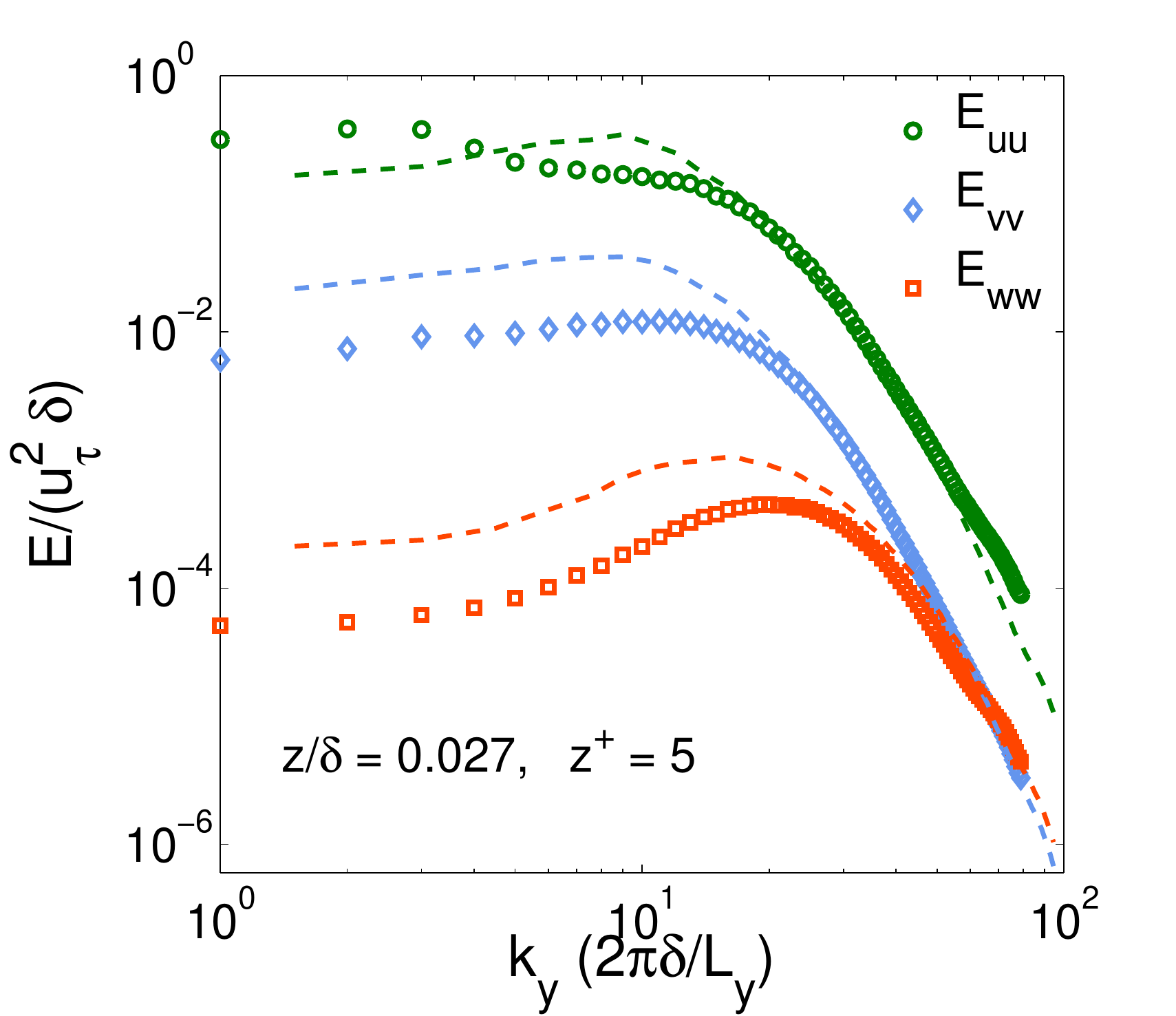} 
\includegraphics[width=0.49\textwidth]{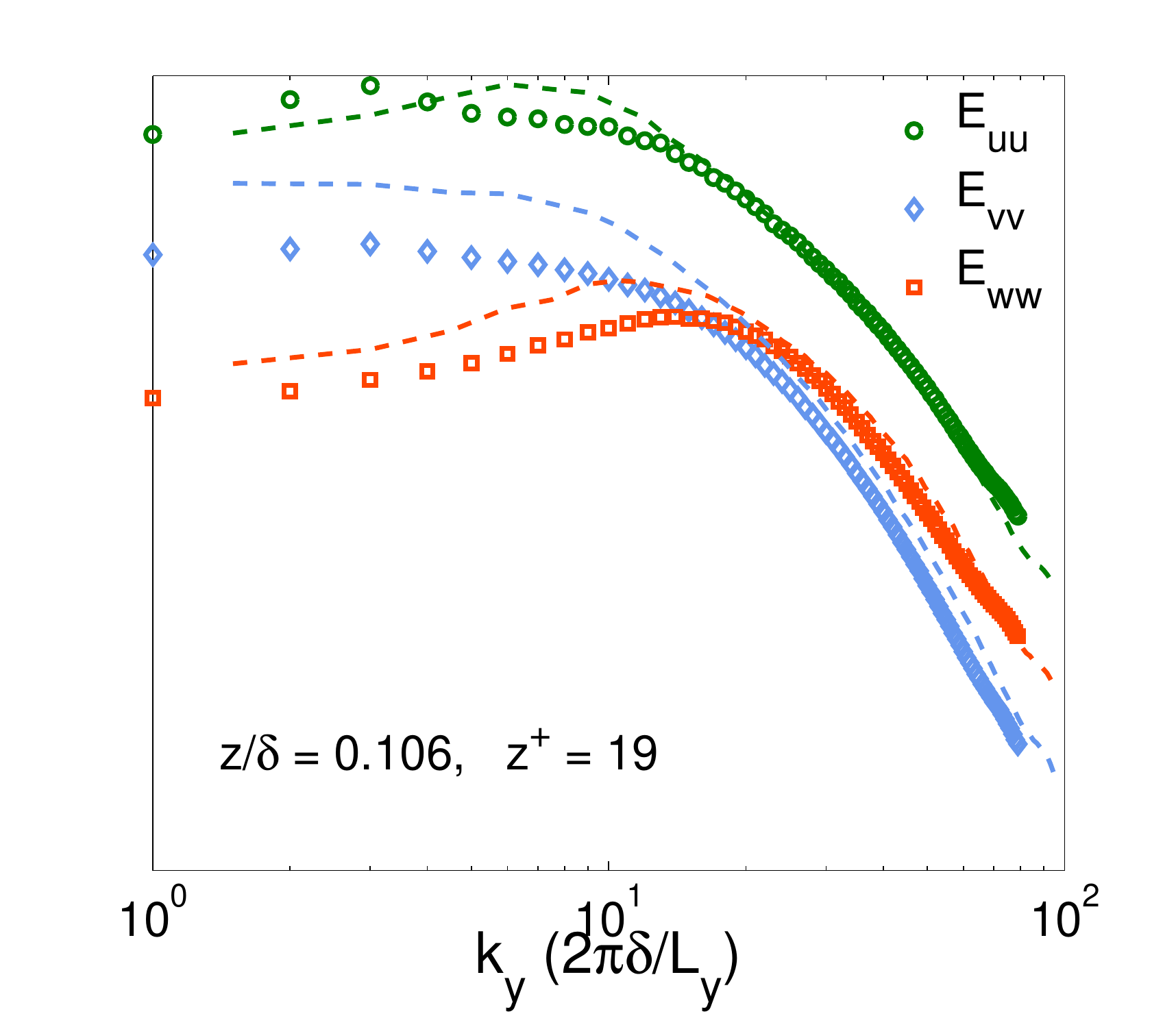}\\ 
(a) \hspace{0.25\linewidth} (b)
\end{center}
\caption{Spanwise energy spectra obtained from band-limited RNL model at Re$_{\tau}=180$, at two wall-normal locations. The RNL system is constrained to a single perturbation wavenumber of $k_x=7$. Dashed lines are DNS data from Moser et al.\cite{Moser99} Symbols are RNL data.}
\label{fig:spectra}
\end{figure}

Finally, we focus on the transverse spatial structure of the fluctuations. Figure \ref{fig:box} already gave an indication that physically realistic structures are generated. More quantitatively, the spanwise spectra can be considered. The RNL case constrained to  the single streamwise-varying wavenumber $k_x = 7$ is shown in Figure \ref{fig:spectra} at two distances from the wall, and compared to DNS. As can be seen, there is  good agreement at small scales. The streamwise velocity spectra from RNL overestimates the DNS spectra at the largest scales while underestimating the peak value which occurs at $k_y\sim 10$, while the spanwise and wall-normal velocity components generated by the RNL simulation underestimate the low wavenumber region of the spectra. Considering the simplicity of the RNL model compared to Navier-Stokes, it can be argued that there is good qualitative agreement in these spectra. 

In summary, we report simulations of a band-limited RNL system which yield improved first- and second-order statistics compared to those obtained from baseline RNL simulations. A single $k_x \neq 0$ ``band" yields mean velocity profiles approaching standard values for the parameters $\kappa$ and $B$. The specific wavenumber to be retained in the model had to be determined  empirically for each of the moderate Reynolds number cases considered here. Increasing the bandwidth to include a set of three adjacent wavenumbers  shows slightly improved statistics when compared with the single $k_x \neq 0$ case at $Re_{\tau}=180$. RNL simulations of the single $k_x \neq 0$ case, however, enable significant savings in computational cost of about a factor of 100 compared to DNS of Navier-Stokes for the Reynolds numbers considered here. These initial results at a range of moderate Reynolds numbers  must be complemented with future simulations at increasing Reynolds numbers to determine whether realistic logarithmic laws  can be achieved at arbitrarily high Reynolds numbers. Also, the asymptotic scaling of the optimal wavenumber and behavior of second-order statistics are topics of continuing investigation.

\begin{acknowledgments}
We thank V. Thomas for helpful discussions on the RNL system. This work is supported by the National Science Foundation (IGERT 0801471, ECCS-1230788 and WINDINSPIRE IIA-1243482).
\end{acknowledgments}

\bibliography{RNL} 

\end{document}